\documentclass{article}
\usepackage{times}
\newcommand{\dd}{\mbox{d}}

\newcommand{\tfrac}[2]{{\textstyle\frac{#1}{#2}}}
\begin{document}
\begin{titlepage}
\title{Bogoliubov's Theory: \\ A Paradigm of Quantum Phase Transitions}
\author{Adriaan M.\ J.\ Schakel\thanks{E-mail:
schakel@physik.fu-berlin.de} \\ National Center for Theoretical Sciences
\\ P.O.\ Box 2-131, Hsinchu, Taiwan 300 \\ and \\ Institut f\"ur
Theoretische Physik \\ Freie Universit\"at Berlin \\ Arnimallee 14,
14195 Berlin }
\date{\today}
\maketitle
\begin{abstract}
This short essay discusses the application of Bogoliubov's theory of
superfluidity in the context of quantum phase transitions. 
\end{abstract}
\end{titlepage}
The importance of N.\ N.\ Bogoliubov's ground-braking paper {\it On the
Theory of Superfluidity} \cite{Bogoliubov} in the development of an
understanding of superfluidity cannot be underestimated \cite{history}.
More than 50 years after the publication of this seminal work, it
continues to play a dominant role in contemporary condensed matter
physics.  It therefore seems appropriate on the occasion of
commemorating Bogoliubov's 90s birthday to submit a short essay
discussing a modern application of his theory in the context of quantum
phase transitions.  Some of the material presented here is more
extensively discussed in the review \cite{Pompo}.  Other recent reviews
can be found in Refs.\ \cite{LG,SGCS,IFT}.

Bogoliubov's theory of superfluidity starts with the Lagrangian
\begin{equation} \label{eff:Lagr}
{\cal L} = \phi^* \bigl[i  \partial_0 - \epsilon(-i  \nabla) +
\mu_0\bigr] \phi - \lambda_0 |\phi|^4,
\end{equation} 
where the complex scalar field $\phi(x)$ describes the atoms of mass $m$
constituting the liquid, $i\partial_0$ is the total energy operator,
while $\epsilon(-i \nabla) = - \nabla^2/2m$ is the kinetic energy
operator, and $\mu_0$ the chemical potential.  The last term with a
positive coupling constant, $\lambda_0 > 0$, represents a weak repulsive
contact interaction.

The theory features a global U(1) symmetry, under which the matter field
acquires an extra phase factor $\phi(x) \rightarrow {\rm e }^{i \alpha}
\phi(x)$, with $\alpha$ the transformation parameter.  Depending on the
ground state, which is determined by the minimum of the potential
energy, the symmetry can be realized in two different ways.  When
$\mu_0<0$, the ground state is at $\phi =0$, and the system is in the
symmetrical state.  As the chemical potential tends to zero, the theory
becomes critical, and when $\mu_0>0$, the global U(1) symmetry is
spontaneously broken by a nontrivial ground state, given by
$|\bar{\phi}|^2 = \mu_0/2\lambda_0$.  This quantity physically denotes
the number density $\bar{n}_0$ of particles residing in the
Bose-Einstein condensate.  

The spectrum of the single-particle excitations in this state is given
by the celebrated Bogoliubov form \cite{Bogoliubov},
\begin{equation}   \label{eff:bogo}
E({\bf k}) = \sqrt{ \epsilon ^2({\bf k}) + 2 \mu_0 \epsilon({\bf k}) },
\end{equation} 
whose most important signature is that at low momentum it takes the
phonon form $E({\bf k}) \sim \sqrt{\mu_0/m} \, |{\bf k}|$ predicted by
Landau.  The spectrum was shown by Beliaev \cite{Beliaev} to remain
gapless when one-loop quantum corrections are included.  And this was
subsequently proven to hold to all orders in perturbation theory by
Hugenholtz and Pines \cite{HP}, meaning that the Bogoliubov theory
describes a gapless mode.  This mode is nothing but the Goldstone mode
accompanying the spontaneous symmetry breakdown of the global U(1)
symmetry, and is the only degree of freedom present in this state.  In
other words, the Bogoliubov theory is a phase-only theory.  At zero
temperature and in the absence of impurities, the phase field is
governed by the effective Lagrangian \cite{effbos}
\begin{equation} \label{eff:Leff}  
{\cal L}_{{\rm eff}} = -\bar{n}\left[\partial_{0}\varphi +
\frac{1}{2m}( {\bf \nabla} \varphi)^{2} \right] + \frac{\bar{n}}{2m
c^{2}}\left[\partial_{0}\varphi + \frac{1}{2m}( {\bf
\nabla}\varphi)^{2}\right]^{2},
\end{equation} 
where ${\bar n}$ is the average particle number density of the system at
rest characterized by a constant phase field $\varphi(x) = {\rm const}$, and
$c$ is the sound velocity, which to a first approximation equals $c =
\sqrt{\mu_0/m}$.  The phase rigidity in the spatial directions, i.e., the
coefficient of $\tfrac{1}{2} (\nabla \varphi)^2$, is seen to be given by
$\bar{n}/m$, while that in the temporal direction by the compressibility
$\kappa$ because
\begin{equation} 
\frac{\bar{n}}{mc^2} = \bar{n}^2 \kappa .
\end{equation} 
Both these rigidities are response functions.  Since the chemical
potential $\mu$ is represented in the effective theory
(\ref{eff:Leff}) by \cite{PWA}
\begin{equation} \label{jo-pwa}
\mu(x) = - \partial_0 \varphi(x),
\end{equation} 
a single differentiation of the effective Lagrangian with respect to
$\mu$ yields the particle number density $n(x) = \bar{n} -(\bar{n}/m
c^{2}) [\partial_{0} \varphi + ( {\bf \nabla} \varphi)^{2}/2 m]$ of
the system slowly varying in space and time
\begin{equation} \label{spaceandtime}
\frac{\partial {\cal L}_{\rm eff}}{\partial \mu(x)} = n(x),
\end{equation} 
while a second differentiation yields the compressibility
\begin{equation} 
\frac{\partial^2 {\cal L}_{\rm eff}}{\partial \mu^2} = \bar{n}^2 \kappa,
\end{equation} 
as required.  It also follows from Eqs.\ (\ref{jo-pwa}) and
(\ref{spaceandtime}) that $n$ and $\phi$ are canonically conjugate
variables \cite{London}.  The form of the effective theory
(\ref{eff:Leff}), especially the combination $\partial_{0}\varphi +
( {\bf \nabla} \varphi)^{2}/2m$ in square brackets is dictated
by Galilei invariance \cite{GWW}.  In cases where this symmetry is
explicitly broken, as in the presence of impurities and at finite
temperature, we expect changes in the relative weights of the
coefficients (see below).

Another, for the further development of the theory of superfluidity
\cite{history}, momentous observation made by Bogolibov was the
so-called depletion of the condensate.  He showed that even at the
absolute zero of temperature not all the particles reside in the
ground state, but \cite{Bogoliubov}
\begin{equation} \label{bec:depl}
\frac{\bar{n}}{\bar{n}_0} -1 \approx \frac{8}{3} \left(\frac{\bar{n}
a^3}{\pi}\right)^{1/2}, 
\end{equation} 
where we replaced the coupling constant with the s-channel scattering
length $a= m \lambda/2 \pi$ \cite{Hugenholtz,AGD}.  (Recall that
$\bar{n}_0$ denotes the density of particles in the condensate.)  Due
to the interparticle repulsion, particles are removed from the
condensate and put in states of finite momentum.  In a strongly
interacting system like superfluid $^4$He, the depletion is such that
no more than about 8\% of the particles condense in the zero-momentum
state \cite{PeOn}.

Despite the depletion of the condensate, the phase rigidity in the
spatial directions was found in Eq.\ (\ref{eff:Leff}) to be given at the
absolute zero of temperature and in the absence of impurities by the
{\it total} average particle number density $\bar{n}/m$.  Since this
coefficient denotes the superfluid particle number density $\rho_{\rm
s}$ (divided by $m^2$), all the particles---not just those residing in
the condensate---participate in the superfluid motion \cite{NoPi}.  This
changes at finite temperature and also when impurities are included:
Galilei invariance is broken then and $\rho_{\rm s}$ no longer equals $m
\bar{n}$.  On the other hand, the phase rigidity in the temporal
direction as well as the first term in the effective Lagrangian
(\ref{eff:Leff}) stay the same.  This is because relation (\ref{jo-pwa})
remains true.  In general we thus have as effective theory \cite{FWGF}
\begin{equation} \label{effprime}
{\cal L}_{{\rm eff}} = -\bar{n} \partial_{0}\varphi -
\frac{\rho_{\rm s}}{2m^2}( {\bf \nabla} \varphi)^{2} + \frac{1}{2}
\bar{n}^2 \kappa (\partial_{0}\varphi)^2 + \cdots .
\end{equation} 

Up to this point we have not specified the external parameter which must
be varied to tune the chemical potential to its critical value where the
system undergoes a phase transition.  In the conventional application of
the Bogoliubov theory, the control parameter is the temperature $T$.
The critical temperature $T_{\rm c}$ can be determined within the theory
by calculating the finite-temperature effective potential and
identifying the temperature at which the minimum starts to shift away
from the origin.  At the one-loop level, one finds \cite{effbos}:
\begin{equation}   \label{bec:jus}
T_{\rm c} = \pi\left[\sqrt{2}\,
\zeta(\tfrac{3}{2})\right]^{-2/3} 
\frac{1}{m} \left( \frac{\mu}{\lambda } \right)^{2/3} - \frac{2}{3}
\frac{ \zeta (\tfrac{1}{2})}{\zeta (\tfrac{3}{2})} \mu ,
\end{equation}
where in obtaining this result a high-temperature expansion has been used.
This is justified because the leading term is of the order $\lambda
^{-2/3}$, which is large for weak-coupling.  Equation (\ref{bec:jus})
expresses the critical temperature in terms of the chemical potential.  From
the experimental point of view, however, it is more realistic to have the
particle number density as independent variable.  One then finds instead
\cite{effbos}:
\begin{equation} \label{deltaT}
\frac{T_{\rm c} - T_0 }{T_0} = c_0 \left(\bar{n} a^3 \right)^\gamma,
\end{equation} 
where we again replaced $\lambda$ with the scattering length $a$, $c_0 =
- \frac{8}{3} \zeta (\tfrac{1}{2})/\zeta (\tfrac{3}{2}) \approx 2.82$,
$\gamma =\tfrac{1}{3}$, and $T_0 = (2 \pi/m)
\left[\bar{n}/\zeta(\tfrac{3}{2})\right]^{2/3}$ is the critical
temperature of a free Bose gas $(\lambda=0)$.  It follows that the
critical temperature is increased by the weak repulsive interaction.
This is qualitatively different from the strongly interacting $^4$He
system.  A free gas with $^4$He parameters at vapor pressure would have
a critical temperature of about 3.1 K, whereas liquid $^4$He becomes
superfluid at the {\it lower} temperature of 2.2 K.  A similar picture
emerges from path-integral Monte Carlo simulations carried out by
Gr\"uter, Ceperley, and Lalo\"e \cite{GCL}.  They found that at low
densities, corresponding to small $a$, the critical temperature is
increased by the repulsive interaction, while at higher densities it is
decreased.  In the weak-coupling limit, they found numerically the same
exponent $\gamma = 0.34 \pm 0.03$ as in Eq.\ (\ref{deltaT}), while the
value of $c_0$ was found to be an order of magnitude smaller: $c_0 =
0.34 \pm 0.06$.  As argued by these authors, a moderate repulsive
interaction suppresses density fluctuations, resulting in a more
homogeneous system.  This facilitates the formation of large so-called
exchange rings necessary to form a Bose-Einstein condensate.  These
exchange rings, as they appear in Feynman's theory of Bose-Einstein
condensation \cite{lambda}, consist of bosons which are cyclicly
permuted in imaginary time (see Ref.\ \cite{StAd} for a recent
account).  At higher densities, the exchange is obstructed because due
to the strong repulsive interaction it is more difficult for the
particles to move.  This leads to a lower critical temperature.

We now turn to the main subject of this essay, and consider the quantum
critical behavior of the Bogoliubov theory first studied by Uzunov
\cite{Uzunov}.  The critical behavior of a system close to a quantum
phase transition is dominated not by thermal fluctuations as in a
classical phase transition at finite temperature, but by quantum
fluctuations.  In this context, the Bogoliubov theory is considered to
be a phenomenological theory similar to the Landau theory of classical
phase transitions.  The system undergoes a quantum transition at the
absolute zero of temperature when the chemical potential approaches a
critical values $\mu_{\rm c}$, which is not necessarily zero as in the
case of the finite-temperature classical transition.  The fine tuning of
the chemical potential can be achieved by varying a number of external
parameters, such as the charge carrier density, the applied magnetic
field, or the impurity strength.  For values of the renormalized
parameter larger than the critical value $\mu>\mu_{\rm c}$, the global
U(1) symmetry is spontaneously broken and the system is superfluid with
a single-particle spectrum given by the gapless Bogoliubov spectrum,
implying that the system is compressible.  On lowering $\mu$, this state
is destroyed and replaced by an insulating state \cite{FWGF}.

In the absence of impurities, the insulating state is a so-called
Mott-insulator, characterized by the absence of phase rigidity in both
spatial and temporal directions, and by an energy gap in the
single-particle spectrum.  This insulating state, which arises solely due
to the repulsive interaction, is consequently incompressible.

On the other hand, in the presence of impurities, the bosons become
trapped by the impurities, i.e., Anderson localized.  The resulting
insulating state is a so-called Bose glass characterized by a
single-particle spectrum that is---like in the superfluid
state---gapless.  This state is therefore also compressible, so that the
compressibility remains finite at the transition.

To account for (quenched) impurities, the following term is added to
the Bogoliubov theory:
\begin{equation} \label{Dirt:dis}
{\cal L}_{\Delta} = \psi({\bf x}) \, |\phi(x)|^2,
\end{equation} 
with $\psi({\bf x})$ a real random field whose distribution is assumed
to be Gaussian \cite{Ma}
\begin{equation}  \label{random}
P(\psi) = \exp \left[-\frac{1}{\Delta_0} \int \dd^d x \, \psi^2({\bf x})
\right],
\end{equation}
and characterized by the impurity strength $\Delta_0$.  Physically, $\psi$
describes impurities randomly distributed in space.  These impurities lead
to an additional depletion of the condensate given in $d$ space dimensions
by \cite{GPS,pla}
\begin{equation}  \label{depDelta}
\bar{n}_\Delta = 2^{d/2-5} \pi^{-d/2}\Gamma(2-d/2) m^{d/2}
\lambda^{d/2-2} \bar{n}_0^{d/2-1} \Delta.
\end{equation} 
The superfluid and normal mass density $\rho_{\rm s}$ and $\rho_{\rm n}$,
respectively now become at the absolute zero of temperature \cite{pla}
\begin{equation} 
\rho_{\rm s} = m\left(\bar{n} -  \frac{4}{d} \bar{n}_\Delta \right), \;\;\;\;
\rho_{\rm n} = \frac{4}{d} m \bar{n}_\Delta.
\end{equation}  
It follows that the normal density is a factor $4/d$ larger than the
mass density $m\bar{n}_\Delta$ knocked out of the condensate by the
impurities.  (For $d=3$ this gives the factor $\tfrac{4}{3}$ first found
in Ref.\ \cite{HuMe}.)  As argued by Huang and Meng \cite{HuMe}, this
implies that part of the zero-momentum states belongs (for $d < 4$) not
to the condensate, but to the normal fluid.  Being trapped by the
impurities, this fraction of the zero-momentum states are localized.  In
other words, the phenomenon of Anderson localization can be accounted
for in the Bogoliubov theory of superfluidity by including a random
field.

The universality class defined by the zero-temperature Bogoliubov theory
is not only relevant to describe the critical behavior of superfluid
films (either with or without impurities), but also to describe that of
other systems, including Josephson junction arrays and superconducting
films.  In the so-called composite-boson limit, where Cooper pairs form
tightly bound states, the BCS theory directly maps onto the Bogoliubov
theory \cite{Haussmann,Pompo}, which is as we argued a phase-only
theory.  But even a weakly interacting BCS system was argued to be in
the same universality class \cite{CFGWY}.  The reason is that the
amplitude fluctuations of the order parameter are not critical at the
transition, not even in the classical superconductor-to-normal
transition in $d=3$ \cite{NgSu}, only the phase fluctuations are.  The
phase of the order parameter therefore constitutes the relevant degree
of freedom, which is precisely the one described by the Bogoliubov
theory.  (See, however, Ref.\ \cite{Ramakrishnan}, where it is argued
that the amplitude fluctuations cannot be neglected, when considering
quantum phase transitions in impure superconducting films.)  The
Bogoliubov theory presumably also forms the basis for the description of
the critical behavior of fractional quantized Hall systems \cite{NP}.

To investigate the role of quantum fluctuations in the Bogoliubov theory
we start with a dimensional analysis.  Since, as far as the quantum
critical behavior of this theory is concerned, the mass $m$ is an
irrelevant parameter, it can be scaled away by introducing $t'=t/m, \;
\mu_0' = m \mu, \; \lambda_0' = \lambda_0 m$.  The engineering dimension
of the various variables is then easily determined as:
\begin{equation}  \label{BT:scale} 
[{\bf x}] = -1, \;\;\;\; [t] = -2, \;\;\;\; [\mu_0] = 2, \;\;\;\;
[\lambda_0] = 2-d,  \;\;\;\; [\phi] = \tfrac{1}{2}d,
\end{equation} 
with $d$ the number of space dimensions, and where we dropped the primes
again.  Note that the time dimension counts double as compared to the
space dimensions.  This is typical for nonrelativistic theories where
the time derivative is accompanied by two space derivatives [see Eq.\
(\ref{eff:Lagr})].  In two space dimensions, the coupling constant
$\lambda_0$ has zero engineering dimension, showing that the
$|\phi|^4$-term is a marginal operator, and that $d_{\rm c}=2$ is the
upper critical space dimension above which the quantum critical behavior
of the Bogoliubov theory becomes Gaussian.  For $d>d_{\rm c}$ quantum
fluctuations are irrelevant, while for $d<d_{\rm c}$ these fluctuations
become crucial.

Let us next compute the one-loop effective potential
\begin{equation} 
{\cal V}_{\rm eff} = - \frac{\mu^2_0}{4 \lambda_0} + \frac{1}{2} \int
\frac{\dd^d k}{(2 \pi)^d} E({\bf k}),
\end{equation} 
with $E({\bf k})$ the gapless Bogoliubov spectrum (\ref{eff:bogo}).  The
integral over the loop momentum yields close to the upper critical dimension
$d=2$:
\begin{equation} \label{epsilon}
{\cal V}_{\rm eff} = - \frac{\mu^2_0}{4 \lambda_0} - \frac{1}{4 \pi 
\epsilon} \frac{m \mu^2_0}{\kappa^\epsilon} + {\cal O}(\epsilon^0),
\end{equation} 
where $\epsilon = 2-d$, and $\kappa$ is an arbitrary renormalization
group scale parameter, with the dimension of an inverse length.  The
right-hand side of Eq.\ (\ref{epsilon}) is seen to diverge when the
upper critical dimension is approached.  The theory can be rendered
ultraviolet finite by introducing a renormalized coupling constant
$\lambda$
\begin{equation} \label{eff:lambdar}
\frac{1}{\hat{\lambda}} = \frac{\kappa^\epsilon}{\lambda_0} + \frac{m}{\pi
\epsilon},
\end{equation}  
where $\hat{\lambda} = \lambda/ \kappa^\epsilon$.  Its definition is
such that for arbitrary $d$, $\hat{\lambda}$ has the same engineering
dimension as $\lambda_0$ in the upper critical dimension $d=2$.  As
renormalization prescription we used the modified minimal subtraction.
The beta function $\beta(\hat{\lambda})$ follows as \cite{Uzunov}
\begin{equation}  \label{beta}
\beta(\hat{\lambda}) = \kappa \left. \frac{\partial \hat{\lambda}}{\partial
\kappa} \right|_{\lambda_0} = -\epsilon \hat{\lambda} + \frac{m}{\pi}
\hat{\lambda}^2.
\end{equation}  
In the upper critical dimension, this yields only one fixed point, viz.\
the infrared-stable (IR) fixed point $\hat{\lambda}^* = 0$.  Below
$d=2$, this fixed point point is shifted to $\hat{\lambda}^* = \epsilon
\pi /m$, implying that the system undergoes a 2nd-order quantum phase
transition.  Above the upper critical dimension, there is no
(nontrivial) renormalization of the coupling constant, which explains why
we omitted the subscript $0$ on $\mu$ and $\lambda$ in Eq.\
(\ref{bec:jus}).

Since Eq.\ (\ref{epsilon}) could be rendered finite solely by a
renormalization of the coupling constant, it follows that the chemical
potential is not renormalized to this order.  As shown by Uzunov these
results remain true to all orders in perturbation theory \cite{Uzunov}.
The reason for this behavior is the special analytic structure of the
nonrelativistic propagator at criticality, representing only particles
propagating forward in time.  As a result, the self-energy (and
consequently $\mu$) is not renormalized and the full 4-point vertex
function is given by a geometric series, leading to the same beta
function (\ref{beta}) found at the one-loop order.  Closely connected to
this is that, despite the nontrivialness of the IR fixed point in $d <
2$, the critical indices characterizing it are Gaussian \cite{Uzunov}.
This conclusion was confirmed by numerical simulations in $d=1$
\cite{BSZ}.

This changes when impurities are included.  A direct application of the
renormalization group \cite{KU} lead to the conclusion that the IR fixed
point becomes instable.  A more careful analysis, using a so-called
double epsilon expansion, shows that the fixed point remains stable upon
including impurities.  The double epsilon expansion was originally
introduced in statistical mechanics by Dorogovtsev \cite{Dorogovtsev} to
treat impurities of finite extend in a classical system.  To
consistently account for these in perturbation theory, one must assume
their dimensionality $\epsilon_{\rm d}$ to be small, and perform in
addition to the usual epsilon expansion, also an expansion in
$\epsilon_{\rm d}$.  The impurities described by Eq.\ (\ref{Dirt:dis})
are static grains which trace out a straight worldlines when time is
included.  In other words, the impurities are line-like in spacetime,
and have also to be treated in a double epsilon expansion, assuming that
their dimensionality $\epsilon_{\rm d}$ is not $1$, but small instead.
The quantum critical behavior of the Bogoliubov theory in $d$ space
dimensions with randomly distributed static impurities tracing out
``worldlines'' of dimensionality $\epsilon_{\rm d}$ falls in the
universality class of a $d$-dimensional classical system with randomly
distributed extended impurities of dimensionality $2 \epsilon_{\rm
d}$---at least to the one-loop order \cite{pla}.  The factor $2$ arises
because, as we mentioned before, in the nonrelativistic Bogoliubov
theory, time dimensions count double as compared to space dimensions.

Besides having a diverging correlation length $\xi$, 2nd-order quantum phase
transitions also have a diverging correlation time $\xi_t$, indicating the
time period over which the system fluctuates coherently.  The way the
diverging correlation time scales with the diverging correlation length,
\begin{equation} \label{zcrit}
\xi_t \sim \xi^z, 
\end{equation} 
defines the so-called dynamic exponent $z$.  The traditional scaling
theory of classical 2nd-order phase transitions is easily extended so as
to include the time dimension \cite{Ma}.  Let $\delta \propto K - K_{\rm
c}$, with $K$ the external control parameter, denote the distance from
the phase transition, so that $\xi \sim |\delta|^{-\nu}$, with $\nu$ the
correlation length exponent.  At the absolute zero of temperature, a
physical observable $O(k_0,|{\bf k}|,K)$ at finite energy $k_0$ and
momentum ${\bf k}$ can in the critical region be written as
\begin{equation} \label{scaling0}
O(k_0,|{\bf k}|,K) = \xi^{d_O} {\cal O}(\xi_t k_0, \xi |{\bf k}|),
\;\;\;\;\;\;\;\; (T=0),
\end{equation} 
where $d_O$ is the scaling dimension of the observable $O$.  The right-hand
side does depend not explicitly on $K$, but only implicitly through $\xi$
and $\xi_t$.

Since a physical system is always at some finite temperature, we have to
investigate how the scaling law (\ref{scaling0}) changes when the
temperature becomes nonzero.  The easiest way to include temperature in a
quantum field theory is to go over to imaginary time $\tau = i t$, with
$\tau$ restricted to the interval $0 \leq \tau \leq \beta$, where
$\beta=1/T$ is the inverse temperature.  The time dimension thus becomes
compactified.  The critical behavior of a phase transition at finite
temperature is still controlled by the quantum critical point provided
$\xi_t < \beta$, so that the system does not notice the finite extend of the
time dimension.  Instead of the zero-temperature scaling (\ref{scaling0}),
we now have the finite-size scaling
\begin{equation} \label{scalingT}
O(k_0,|{\bf k}|,K,\beta) = \beta^{d_O/z} {\cal O}(\beta k_0 ,
\beta^{1/z} |{\bf k}|,\beta/\xi_t), \;\;\;\;\;\;\;\; (T \neq 0).
\end{equation} 
The distance to the quantum critical point is measured by the ratio
$\beta/\xi_t \sim |\delta|^{z\nu}/T$.

Let us apply these general considerations to the effective theory
(\ref{effprime}) \cite{FF,FWGF}.  The singular part of the free energy
density $f_{\rm sing}$, which scales near the transition as
\begin{equation} 
f_{\rm sing} \sim \xi^{-(d+z)},
\end{equation} 
arises from the low-energy, long-wavelength fluctuations of the Goldstone
field.  The ensemble averages give
\begin{equation} 
\langle (\nabla \varphi)^2 \rangle \sim \xi^{-2}, \;\;\;\;
\langle (\partial_0 \varphi)^2 \rangle \sim \xi_t^{-2} \sim \xi^{-2z} .
\end{equation} 
Combined, these hyperscaling arguments yield the following scaling of
the rigidity constants:
\begin{equation} \label{hyperrho} 
\rho_{\rm s} \sim \xi^{-(d+z-2)}, \;\;\;\; \bar{n}^2 \kappa \sim
\xi^{-(d-z)} \sim |\delta|^{(d-z)\nu}.
\end{equation} 
The first conclusion is consistent with the universal jump in the superfluid
density predicted by Nelson and Kosterlitz \cite{NeKo} for a
Kosterlitz-Thouless phase transition which corresponds to taking $z=0$ and
$d=2$.

In an impure system undergoing an Anderson transition, the
compressibility $\bar{n}^2 \kappa$ is nonsingular at the critical point
and hence $z=d$ for repulsively interacting bosons in an impure media
\cite{FF}.  Surprisingly, the same conclusion holds for an impure {\it
fermionic} system \cite{IFT}.  For $d=1$ it follows that space and time
appear symmetric as in a relativistic theory.

In a clean system, on the other hand, with a density-driven Mott
transition, i.e., $\delta \propto \mu - \mu_{\rm c}$, $f_{\rm sing}$
can also be directly differentiated with respect to the chemical
potential to yield for the singular part of the compressibility
\begin{equation} 
\bar{n}^2 \kappa_{\rm sing} \sim |\delta|^{(d+z)\nu -2}. 
\end{equation} 
In this case $\bar{n}^2\kappa \sim \bar{n}^2 \kappa_{\rm sing}$, so that
$z \nu =1$ \cite{FWGF} in accord with the Gaussian values $\nu=\tfrac{1}{2}, \; z=2$
found by Uzunov \cite{Uzunov} for the pure case in $d < 2$.

The above hyperscaling arguments have been extended by Fisher,
Grinstein, and Girvin \cite{FGG} to include a $1/|{\bf x}|$-Coulomb
potential.  This potential is important for the quantum phase
transitions in charged systems because the Coulomb repulsion suppresses
fluctuations in the charge density and simultaneously enhances those in
the canonically conjugate variable $\phi$, thereby disordering the
ordered state.  The quadratic terms of the effective theory in Fourier
space become when the $1/|{\bf x}|$-Coulomb potential is included
\cite{FGG}
\begin{equation} 
{\cal L}_{\rm eff}^{(2)} = \frac{1}{2} \left(\rho_{\rm s} {\bf k}^2 -
\frac{1}{\hat{e}^2} k_0^2 \, |{\bf k}|^{d-1} \right) |\varphi(k_0,{\bf
k})|^2,
\end{equation}  
where $\hat{e}$ is the renormalized charge.  Using similar hyperscaling
arguments as before, one finds that this charge scales as
\begin{equation} 
\hat{e}^2 \sim \xi^{1-z}.
\end{equation} 
Arguing  that  in  the  presence  of random  impurities  the  charge  is
nonsingular at the transition, the authors of Ref.\ \cite{FGG} concluded
that 
\begin{equation} \label{z}
z=1.
\end{equation}    
This again is an exact result which replaces the value $z=d$ of the
neutral system in an impure media.

Most experiments on quantum phase transitions in charged systems measure
the conductivity $\sigma$.  To describe such type of systems, we
minimally couple the Bogoliubov theory to an electromagnetic vector
potential $(A_0,{\bf A})$.  The conductivity turns out to be related to
the superfluid mass density via \cite{CFGWY}
\begin{equation} \label{conductivity}
\sigma(k) = i \left(\frac{e}{m}\right)^2 \frac{\rho_{\rm s}(k)}{k_0}.
\end{equation} 
On account of the scaling relation (\ref{hyperrho}), it then follows
that
\begin{equation} 
\sigma \sim \xi^{-(d-2)},
\end{equation} 
implying that the conductivity and therefore the resistivity is a
marginal operator in two space dimensions \cite{Wegner}.  

The magnetic field $H$ scales with $\xi$ as $H \sim \Phi_0/\xi^2$, where
$\Phi_0=2 \pi/e$ is the magnetic flux quantum.  This implies that the
scaling dimension $d_{\bf A}$ of ${\bf A}$ is unity,
\begin{equation} 
d_{\bf A} = 1,
\end{equation} 
so that $|{\bf A}| \sim \xi^{-1}$.  From this it in turn follows that
the electric field $E=|{\bf E}|$ scales as $E \sim \xi_t^{-1} \xi^{-1}
\sim \xi^{-(z+1)}$, and that the scaling dimension $d_{A_0}$ of $A_0$ is
$z$,
\begin{equation} 
d_{A_0} = z,
\end{equation} 
so that $A_0 \sim \xi_t^{-1} \sim \xi^{-z}$.  

Let us now be specific and consider quantum phase transitions triggered
by changing either the applied magnetic field, i.e., $\delta \propto H -
H_{\rm c}$, or the charge carrier density, i.e., $\delta \propto n -
n_{\rm c}$.  For DC ($k_0=0$) conductivities in the presence of an
external electric field $E$ we have on account of the general
finite-size scaling form (\ref{scalingT}) with $k_0=|{\bf k}|=0$:
\begin{equation}  \label{scalingE}
\sigma(K,T,E) = \varsigma(|\delta|^{\nu z}/T,|\delta|^{\nu (z+1)}/E).
\end{equation}
This shows that conductivity measurements close to a quantum critical
point of the kind discussed here should in general collapse onto two
branches when plotted as function of the dimensionless combinations
$|\delta|^{\nu z}/T$ and $|\delta|^{\nu (z+1)}/E$: a lower branch
bending down for the insulating state and an upper branch tending to
infinity for the other state.  The best collapse of the data determines
the values of $\nu z$ and $\nu (z+1)$.  In other words, the temperature
and electric-field dependence determine the critical exponents $\nu$ and
$z$ independently.  

The table below shows experimental data for the critical exponents $z$
and $\nu$ of the superconductor-to-insulator transition in thin films,
the Hall-liquid-to-insulator transition in fractional quantized Hall
systems, and the conductor-to-insulator transition in silicon MOSFET's
at extremely low electron number densities.  \\[.4cm]

\begin{tabular}{c|cc} 
Transition & $z$ & $\nu$ \\ \hline \\[-.2cm]
Superconductor-to-Insulator \cite{HPsu1,YaKa} & $1.0 \pm 0.1$ & $1.36 \pm
0.05$ \\[.2cm] 
Hall-Liquid-to-Insulator \cite{WTPP,WET} & $\approx 1.0$ & $\approx 2.3$
\\[.2cm] 
Conductor-to-Insulator \cite{PFW,KSSMF} & $0.8 \pm 0.1$ & $1.5 \pm 0.1$
\\[.2cm] \hline 
\end{tabular}
\\[.7cm] 

A few remarks are in order.  First, the values for the dynamic exponent
$z$ found in these systems are in accordance with the prediction $z=1$
recorded in Eq.\ (\ref{z}), which was obtained using general
hyperscaling arguments for an impure system with a $1/|{\bf x}|$-Coulomb
potential.  Second, the values of the critical exponents characterizing
the Hall-liquid-to-insulator transition are universal and independent of
the filling factor---whether an integer or a fraction.  Third, earlier
experiments on silicon MOSFET's at lower densities seemed to confirm the
general believe, based on the work by Abrahams {\it et al.} \cite{AALR},
that such two-dimensional electron systems do not undergo a quantum
phase transition.  In that paper, where electron-electron interactions
were ignored, it was demonstrated that impurities always localize the
electrons at the absolute zero of temperature, thus excluding conducting
behavior.  Apparently, the situation changes drastically at low electron
number densities, where the $1/|{\bf x}|$-Coulomb interaction becomes
important.  The values of the critical exponents found for this
transition are surprisingly close to those found for the
superconductor-to-insulator transition.  Since further experiments in an
applied magnetic field \cite{SKS} also revealed a behavior closely
resembling that near the superconductor-to-insulator transition, it is
speculated that the conducting state in silicon MOSFET's is in fact
superconducting.  \\

\noindent
{\bf Acknowledgments}

I'm grateful to NCTS and B. Rosenstein for the financial support and the
hospitality at the Center in Hsinchu, Taiwan, where this work was
completed.
\end{document}